\documentclass[12pt]{iopart}

\usepackage{ulem}
\usepackage{color}
\usepackage{graphicx}
\usepackage{epsfig}
\usepackage{bm}

\def\nn {\nonumber}

\newcommand{\be}{\begin{equation}}
\newcommand{\ee}{\end{equation}}
\newcommand{\bea}{\begin{eqnarray}}
\newcommand{\eea}{\end{eqnarray}}

\newcommand{\ep}{\epsilon}
\newcommand{\om}{\omega}
\newcommand{\vk}{\vec k}

\newcommand{\del}{\partial}

\begin{document}

\title[]{Finite size effect of hadronic matter on its transport coefficients}

\author{Subhasis Samanta$^{1,a}$, Sabyasachi Ghosh$^{2,3,b}$, Bedangadas Mohanty$^{1,c}$}

\address{$^1$School of Physical Sciences, National Institute of Science 
Education and Research, Bhubaneswar, HBNI, Jatni, 752050, India}
\address{$^2$ Indian Institute of Technology Bhilai, GEC Campus, Sejbahar, Raipur-492015, 
Chhattisgarh, India}
\address{$^3$Department of Physics, University of Calcutta, 92, 
A. P. C. Road, Kolkata - 700009, India}

\ead{$^a$subhasis.samant@gmail.com, $^b$sabyaphy@gmail.com, $^c$bedanga@niser.ac.in}

\vspace{10pt}

\begin{abstract}
We have theoretically investigated the finite system size effect of hadronic matter on its
transport coefficients like shear viscosity, bulk viscosity,
and electrical conductivity. We have used a Hadron Resonance Gas (HRG) model to calculate the 
thermodynamical quantities like entropy density, speed of sound
and also the above transport coefficients. All these quantities are found to be sensitive to
finite system size effects of hadronic matter.
The effect of finite system size is found to be more when the system is
at low temperatures and gets reduced at high temperatures. Owing to the intimate
linking between system size and centrality, we have presented the centrality
dependence of transport coefficients. We have also explored to link of our results 
with the macroscopic picture of hydrodynamical evolution.  
\end{abstract}

%
%
%
%
%

\section{Introduction}
\label{sec:intro}

The research on estimation of transport coefficients for the medium,
produced in high energy heavy ion collision experiments, received a considerable attention in the 
scientific community, when RHIC experiments announced that they have got
a QCD medium, having a very small value of shear viscosity to entropy density
($\eta/s$) ratio \cite{Romatschke:2007mq,Luzum:2008cw,Roy:2012jb}. 
%
%
In high energy
nuclear physics, the topic becomes more exciting when we notice that the traditional
QCD theory~\cite{Arnold} predicts 10-20 times larger value of $\eta/s$ than the experimental value
or the lower bound~\cite{KSS}. As an alternative treatments of QCD theory, different effective
QCD models~\cite{Purnendu,Redlich_NPA,Marty,G_CAPSS,Weise2,Kinkar,G_IFT,Deb,Tawfik}
and hadronic models~\cite{ Itakura,Dobado,Nicola,Weise1,SSS,Ghosh_piN,Gorenstein,HM,Kadam:2015xsa,Hostler} 
have attempted to estimate this $\eta/s$ in recent times. Some attempts are also made by using 
some transport simulations~\cite{Bass,Muronga,Plumari,Pal}, where Kubo-type correlators
are generated for estimating shear viscosity.
Some estimations are also done from Lattice QCD calculations~\cite{Meyer_eta,LQCD_eta2}.
%
Other transport coefficients
like bulk viscosity and electrical conductivity are also
important to estimate from the same dynamical framework 
for the strongly interacting medium. 
Some of the work in the literature for the microscopic
calculations of bulk viscosity can be found in the Refs.~\cite{{Paech1},{Gavin},{Arnold_bulk},{Prakash},{Tuchin},
{Tuchin2},{Nicola},{Marty},{De-Fu},{Redlich_NPA},{Redlich_PRC},{G_IFT},{Deb},{Tawfik},
{Purnendu},{Santosh},{Sarkar},{HM},{Kadam:2015xsa},{Kadam:2015fza},
{Sarwar:2015irq},{Hostler},{Nicola_PRL},{SG_NISER},{Meyer_zeta},{Dobado_zeta1},{Dobado_zeta2},{Saha:2015lla}},
and those for electrical conductivity can be found in the Refs.~\cite{LQCD_Ding,LQCD_Arts_2007,LQCD_Buividovich,
LQCD_Burnier,LQCD_Gupta,LQCD_Barndt,LQCD_Amato,Cassing,Puglisi,Greif,
{Marty},{PKS},{Finazzo},{Lee},{Nicola_PRD},{Greif2},{Ghosh:2016yvt}}.

Here our interest is on finite volume effect in transport coefficient calculations, since
the matter created due to the energy deposition of the colliding nuclei has a finite volume.
The volume of the system depends on the size of the colliding nuclei, the center of mass energy
and centrality of the collision.
Different effects of finiteness of the system size have been discussed in the literature
\cite{Luscher:1985dn, Elze:1986db, Gasser:1987ah, Spieles:1997ab,
Gopie:1998qn, Kiriyama:2002xy, Abreu:2006pt, Shao:2006gz, Yasui:2006qc, 
Palhares:2009tf, Luecker:2009bs, Fraga:2011hi,
Abreu:2011rj, Abreu:2011zzc, Bhattacharyya:2012rp,
Bhattacharyya:2014uxa, Bhattacharyya:2015zka, Bhattacharyya:2015kda, Magdy:2015eda,
Redlich:2016vvb,RedlichV, Xu:2016skm,Bhattacharyya:2015pra,Nachiketa1,Nachiketa2,KinkarV},
where major investigations~\cite{Spieles:1997ab,Kiriyama:2002xy,Abreu:2006pt,Shao:2006gz, Yasui:2006qc,
Palhares:2009tf,Bhattacharyya:2012rp,Bhattacharyya:2014uxa, Bhattacharyya:2015zka, Bhattacharyya:2015kda,Magdy:2015eda} 
are focus to see the volume effect in quark-hadron phase transition.
The present article is aimed to observe a finite system size effect of the medium on the 
estimation of transport coefficients. Specifically we discuss the relative
change in the values of the transport coefficients because of transformation from
infinite matter to finite matter. 

In a classical
view we may not expect different values of transport
coefficient for a fluid having different system size.
However, for fluid where the system size is small enough for quantum 
effect to play a role, one may
expect a possibility of finite system size effect. We have tried to explore this fact for the tiny
QCD matter, produced in the high energy heavy ion collision experiment. We realize the lower bound of
shear viscosity to entropy density ratio ($\eta/s$) arises due to lowest possible quantum fluctuation
of fluid, which can never be ignored even in the infinite coupling limit. 
However, in classical view, one can easily think about $\eta/s\rightarrow 0$ in 
this infinite coupling limit.
Since we know that
the $\eta/s$ of RHIC matter is surprisingly close to this lower bound \cite{Song:2010mg,Schenke:2010rr},
therefore, we may associate this matter with the lowest possible quantum fluctuation and we
may also consider other possible quantum effect coming from the finite system size effect.
Following the earlier
Refs.~\cite{Bhattacharyya:2012rp,Bhattacharyya:2014uxa,Bhattacharyya:2015zka,Bhattacharyya:2015kda,Redlich:2016vvb,RedlichV}
a finite lower momentum cut-off has been adopted to incorporate the 
finite system size picture of the medium and we have applied it in different transport coefficients
calculations. 

The article is organized as follows. Next section has covered the brief formalism part
of transport coefficients and finite system size picture of HRG model. Then we have analyzed our numerical
output in the result section and at last, we summarize our work.
 
\section{Formalism}
\label{sec:formal}
\subsection{transport coefficients}
In this work, we calculate the above transport coefficients within the 
hadron gas resonance (HRG) model, so we have to add the contributions 
of all mesons ($M$) and baryons ($B$) for getting total transport coefficients
of hadronic matter. The mathematical structure of transport coefficients,
obtained from the one-loop diagram in quasi-particle Kubo approach and 
relaxation time approximation (RTA) in kinetic theory approach, are exactly same,
therefore, we start with the standard expressions of $\eta$~\cite{Nicola,Weise1,G_IJMPA}, 
$\zeta$~\cite{Nicola,G_IFT}
and $\sigma$~\cite{Nicola_PRD,Nicola,Ghosh:2016yvt}: 
\bea
\eta&=&\sum_{B}\frac{g_{B}}{15 T}\int \frac{d^3\vk}{(2\pi)^3}
\tau_{B}\left(\frac{\vk^2}{\om_{B}}\right)^2[n^{+}_{B}(1-n^{+}_{B})
\nn\\
&& + n^{-}_{B}(1-n^{-}_{B})]+\sum_{M}\frac{g_{M}}{15 T}\int \frac{d^3\vk}{(2\pi)^3}
\tau_{M}\left(\frac{\vk^2}{\om_{M}}\right)^2
\nn\\
&&~~~~~~~~~~~~~~~~~~~~~~~~~~~~n_{M}(1+n_{M})~;
\label{eta_G}
\eea
\bea
\zeta&=&\sum_{B}\frac{g_{B}}{T}\int \frac{d^3\vk}{(2\pi)^3\om_{B}^2}
\tau_{B}\left\{\left(\frac{1}{3}-c_s^2\right)\vk^2 
\right.\nn\\
&&\left.- c_s^2 m_B^2\right\}^2\left[ n^{+}_{B}\Big(1-n^{+}_{B}\Big ) 
+ n^{-}_{B}\Big (1-n^{-}_{B}\Big )\right]
\nn\\
&&+\sum_{M}\frac{g_{M}}{T}\int \frac{d^3\vk}{(2\pi)^3\om_{M}^2}
\tau_{M}\left\{\left(\frac{1}{3}-c_s^2\right)\vk^2 
\right.\nn\\
&&\left.
- c_s^2 m_M^2 \right\}^2n_{M}(1+n_{M});
\label{zeta_Gmu0}
\eea
%
%
%
\bea
\sigma&=&\sum_{B}\frac{g_{B}e_B^2}{3 T}\int \frac{d^3\vk}{(2\pi)^3}
\tau_{B}\left(\frac{\vk}{\om_{B}}\right)^2[n^{+}_{B}(1-n^{+}_{B})
\nn\\
&&~~~~~~~~~~~~~~~~~~~~~ + n^{-}_{B}(1-n^{-}_{B})]
\nn\\
&&+\sum_{M}\frac{g_{Me_M^2}}{3 T}\int \frac{d^3\vk}{(2\pi)^3}
\tau_{M}\left(\frac{\vk}{\om_{M}}\right)^2n_{M}(1+n_{M})~,
\nn\\
\label{sigma_G}
\eea
where $g_{B}$ and $g_{M}$ are degeneracy factors for baryons (fermions) $B$ 
 and mesons (bosons) $M$ respectively.
The $n_B^{\pm}$ ($\pm$ stand for particle and anti-particle respectively)
is Fermi-Dirac (FD) distribution function of $B$ having 
energy $\om_B=\{\vk^2 +m_B^2\}^{1/2}$ and $n_M$ is Bose-Einstein (BE) distribution
function of $M$ with energy $\om_M=\{\vk^2 +m_M^2\}^{1/2}$, 
where $m_B$ and $m_M$ denote the masses of B and M; $\vk$
stands for their momentum.
During calculation of
$\sigma$, we have to take care of isospin degeneracy factors of charged hadrons only.
More explicitly, we can write the charge factors of baryons and mesons as 
$e_B^2=(N^e_B e)^2$ and $e_M^2=(N^e_M e)^2$ with $e^2=4\pi/137$, where $N^e_B$ and $N^e_M$ are
their electrical charge number in units of electrons; e.g. for $\Delta^{++}$ it is $N^e_B=2$.

\subsection{HRG Model}
In the HRG model, the system consists of all
the hadrons and resonances. Different versions of HRG
model may be found in Refs.
\cite{Hagedorn:1980kb, Rischke:1991ke, Cleymans:1992jz,
BraunMunzinger:1994xr, Cleymans:1996cd, Yen:1997rv,
BraunMunzinger:1999qy, Cleymans:1999st, BraunMunzinger:2001ip,
BraunMunzinger:2003zd, Karsch:2003zq, Tawfik:2004sw, Becattini:2005xt,
Andronic:2005yp, Andronic:2008gu,Begun:2012rf, Andronic:2012ut,
Tiwari:2011km, Fu:2013gga, Tawfik:2013eua,
Bhattacharyya:2013oya, Bhattacharyya:2015zka,Kadam:2015xsa,
Kadam:2015fza, Albright:2014gva, Albright:2015uua,
Begun:2016cva,Adak:2016jtk, Xu:2016skm,Bhattacharyya:2015pra}.
Here the transport properties of hadronic matter with
a finite volume is studied by using the simplest version namely the
ideal or non-interacting HRG model.
The grand canonical partition function of a hadron resonance 
gas~\cite{BraunMunzinger:2003zd, Andronic:2012ut} can be written as,
\begin {equation}
 \ln Z^{id}=\sum_H \ln Z_H^{id},
\end{equation}
where sum is over all the hadrons, $id$ refers to ideal {\it i.e.},
non-interacting HRG. For individual hadron $H$,
\bea
\ln Z_H^{id}&=&\pm \frac{Vg_H}{(2\pi)^3}\int_0^\infty d^3\vk \ln[1\pm\exp(-(\om_H-\mu_H)/T)]
\nn\\
&=&\pm \frac{Vg_H}{2\pi^2}\int_0^\infty {|\vk|}^2 d|\vk|\ln[1\pm\exp(-(\om_H-\mu_H)/T)],
\nn\\
\eea
where $\om_H=\sqrt{{|\vk|}^2+m^2_H}$ is the single particle energy, $m_H$ is
the mass of the hadron, $V$ is the volume of the system, $g_H$ is the degeneracy
factor for the hadron and $T$ is the temperature of the system. 
In the above expression
$\mu_H=B_H\mu_B+S_H\mu_S+Q_H\mu_Q$ is the chemical potential and
$B_H,S_H,Q_H$ are respectively the baryon number, strangeness and charge
of the hadron, $\mu^,s$ being corresponding chemical potentials.
The upper and lower sign of $\pm$ corresponds to baryons and mesons respectively.
In this work we have incorporated all the hadrons listed in
the particle data book \cite{Olive:2016xmw} up to mass of 3 GeV.
 
Now let us come to the finite volume picture. The constituents in the medium
are restricted to move within the volume of the medium. If the volume is too 
small to ignore the quantum effect then one can roughly relate the system with
the quantum mechanical picture of particle in a box. Unlike to classical
picture, the lowest possible momentum or energy of the constituents will
not be zero. There will be lower momentum cut-off or zero point energy,
below which constituents never go.
We have considered the lower momentum cut-off
$k_{min}=\pi/R=\lambda$(say), where $R$ is the size of a cubic volume 
$V\sim R^3$ \cite{Bhattacharyya:2012rp, Bhattacharyya:2014uxa, Bhattacharyya:2015zka} of
the system.
With this cut-off the partition function of individual hadron $H$ becomes
\begin{equation}\label{eq:zi}
 \ln Z_H^{id}=\pm \frac{Vg_H}{2\pi^2}\int_\lambda^\infty {|\vk|}^2 d|\vk|
 \ln[1\pm\exp(-(\om_H-\mu_H)/T)].
\end{equation}
In principle one should sum over discrete momentum values but for
simplicity we integrate over continuous values of momentum.
This simplified picture of finite size effect by 
implementing lower momentum cut-off is justified in Refs.~\cite{Redlich:2016vvb,RedlichV}.
It is nicely demonstrated in Fig.~(1) of Ref.~\cite{Redlich:2016vvb}.
Along with the discretization of momentum, surface and curvature effect may
appear in the finite size picture~\cite{Shao:2006gz} but we don't consider
these effect in the present work for simplicity.

. 

From partition function we can calculate various thermodynamic
quantities of interest. The partial pressure $P$,
the energy density $\ep$  
can be calculated using the standard definitions,
\bea
P&=&\sum_{H=M,B}T\frac{\partial \ln Z_H^{id}}{\partial V}
\nn\\
\ep&=& -\frac{1}{V} \sum_{H=M,B} \left(\frac{\partial \ln Z_H^{id}} {\partial\frac{1}{T}}\right)_{\frac{\mu}{T}}
\nn\\
&=&\sum_{H=M,B} \frac{g_H}{2\pi^2}\int_\lambda^\infty \frac{{|\vk|}^2 d|\vk|}{\exp[(\om_H-\mu_H)/T]\pm1}\om_H,
\nn\\
\eea

With help of $P$, $\ep$, we can get entropy density (for $\mu_H=0$) as
\be
s=\frac{P+\ep}{T}~.
\label{s_HRG}
\ee
From these thermodynamical quantities, one can find the square of speed
of sound as
\be
c_s^2=\left(\frac{\del P}{\del \ep}\right)=\frac{s}{T\frac{ds}{dT}}~.
\ee
\section{Numerical results and discussion}
\label{sec:num}
\begin{figure}
\centering
\includegraphics[width=0.5 \textwidth]{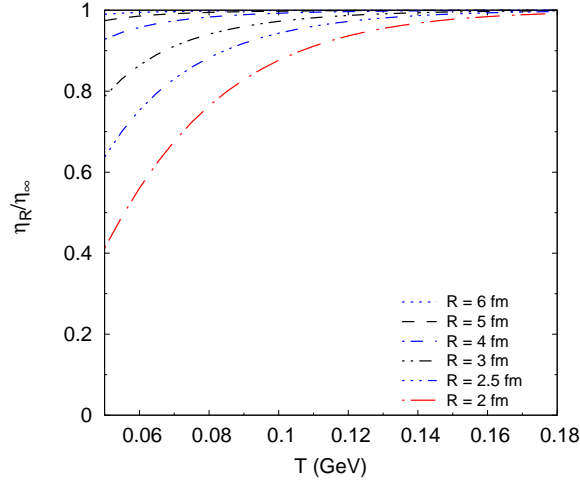}
\caption{(Color online) Temperature dependence of shear viscosity with finite 
size effect ($\eta_R$), normalized by without finite system size effect ($\eta_\infty$) at $\mu = 0$.
This normalized quantity is calculated for different values of finite system size ($R$) in HRG model.}
\label{fig:eta_relative_T}
\end{figure}
\begin{figure} 
\begin{center}
\includegraphics[width=0.42 \textwidth]{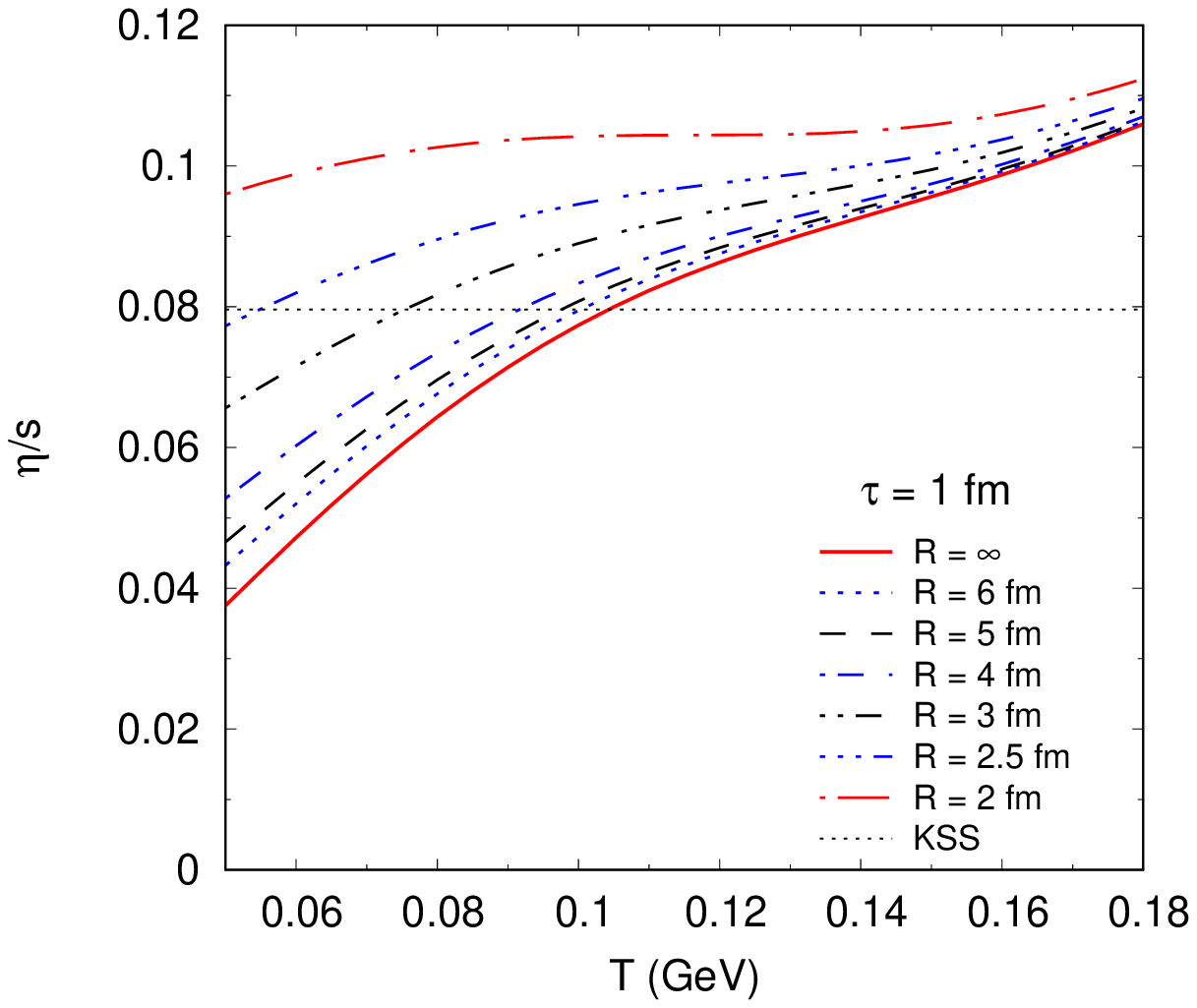}
\includegraphics[width=0.42 \textwidth]{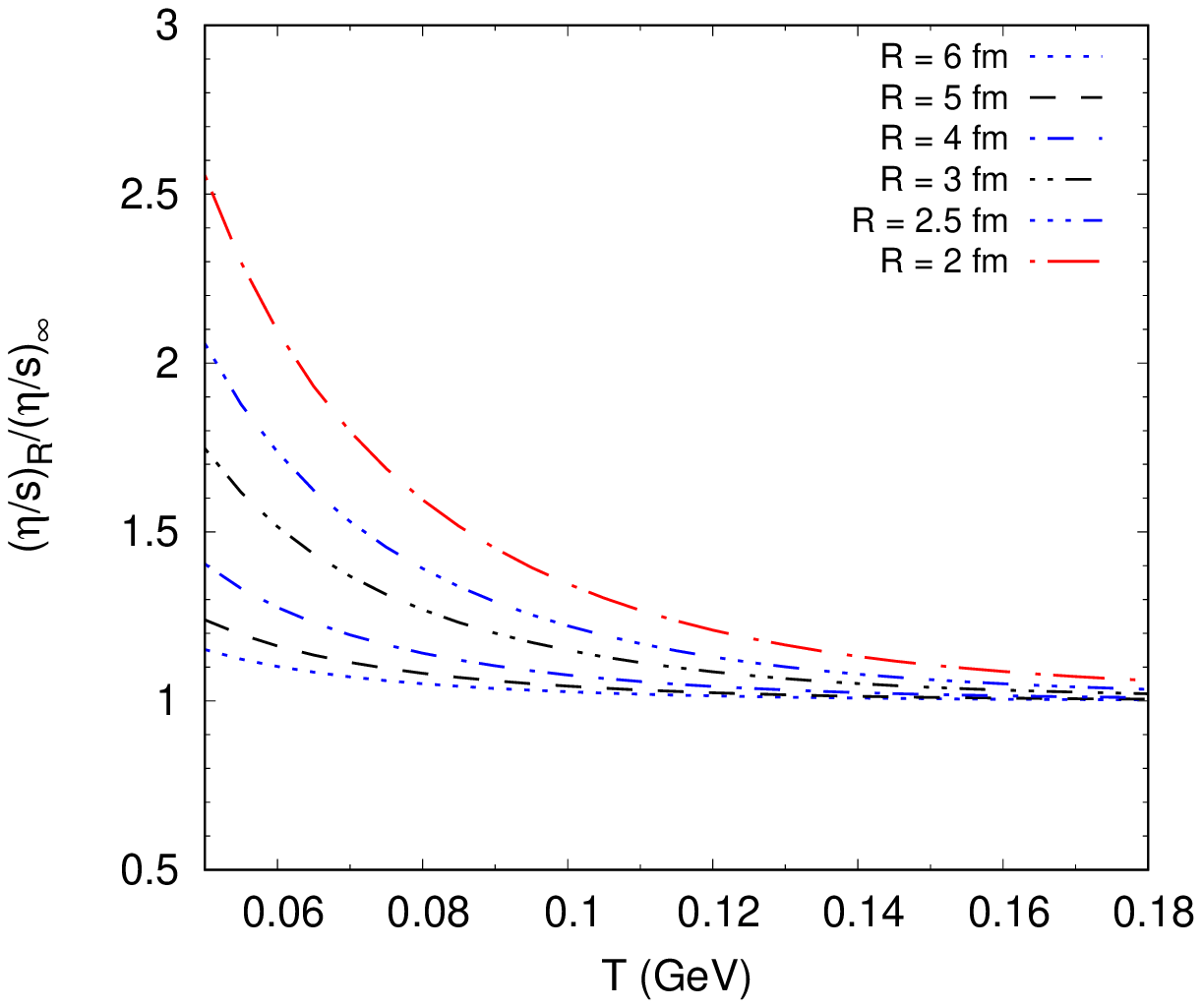}
\caption{(Color online) Left: Shear viscosity to entropy density ratio is plotted versus $T$
for different values of $R$. The viscosity calculations are done for
fixed value of relaxation time $\tau=1$ fm
and $\mu = 0$. Horizontal dotted line stands for KSS bound.
Right: Fractional change in $\eta/s$ at different values of $R$
with respect the infinite matter result.} 
\label{fig:eta_s_T}
\end{center}
\end{figure}
%
The transport coefficients are calculated for two cases.
In one case, they are calculated by considering
finite system size effect and in other case, without considering such an 
effect.
In former case, the limit of integration in Eqs.~(\ref{eta_G}), (\ref{zeta_Gmu0}),
(\ref{sigma_G})
is $\lambda$ to $\infty$, while in latter case, the standard thermodynamical limit
from $0$ to $\infty$ is taken.
Dividing former by latter, we will get ratios of
different transport coefficients, which may provide us an insight about
finite system size effect on different transport coefficients. 
The ratios for $\eta$, $\zeta$ and $\sigma$ are shown in
Figs.~\ref{fig:eta_relative_T}, \ref{fig:bulk_relative_T} and \ref{fig:el_relative_T}
and the related discussions are presented below.

Let us first come to the shear viscosity case, shown in Fig.~\ref{fig:eta_relative_T}, 
where the ratios $\eta_R/\eta_{\infty}$ are plotted versus $T$. Here, 
$\eta_R$ denotes the shear viscosity of hadronic matter, having a system volume with radius $R$,
while $\eta_{\infty}$ stands for shear viscosity of infinite hadronic matter.
In other words, the calculations of $\eta_R$ and $\eta_{\infty}$ are 
the shear viscosity calculations with and without considering finite system size effect, respectively.
We have taken six different sizes of $R = 2,~ 2.5,~ 3,~ 4,~ 5,~ 6$ fm, which
are motivated from the values of $R$ extracted by a HRG model fits to particle
yields in high energy heavy ion collision experiment~\cite{STAR_17}.
Fig.~\ref{fig:eta_relative_T} shows that the $\eta_R/\eta_\infty$ ratio decreases when 
values of $R$ decreases. As momentum cut-off $k_{\rm min}=\pi/R$ becomes larger for smaller
values of $R$, hence thermodynamical phase space shrinks as $R$ decreases. This is the 
reason for which shear viscosity $\eta$ reduces by lowering the $R$.
$\eta$ also decreases with decrease in $T$ of the system. The reason is as follows.
The thermal distribution probability of medium constituents at lower momentum
is enhanced when we reduce the temperature. Now, by introducing lower momentum
cut-off, we are subtracting the statistical weight of low momentum zone. When
we go from high temperature to low temperature domain, this subtraction becomes
larger and therefore, we are getting smaller $\eta$ at small temperature zone.
The ratio $\eta_R/\eta_{\infty}$ approaches towards unity at high $R$ and $T$.
It means that the statistical weight, subtracted by lower momentum cut-off, is
quite small when one approaches toward high $R$ and $T$ zone.
%
%

\begin{figure}
\centering
\includegraphics[width=0.5 \textwidth]{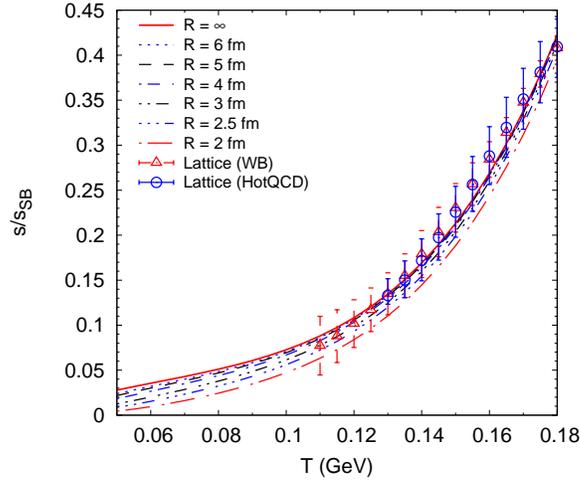}
\caption{(Color online) The ratio of entropy density to its Stefan-Boltzmann (SB) 
value ($s/s_{SB}$) is 
plotted versus $T$ for different values of $R$. Lattice QCD data from the 
WB group~\cite{Borsanyi:2013bia} (triangles)
and Hot QCD group~\cite{Bazavov:2014pvz} (circles) are added.}
\label{fig:s_sSB_T}
\end{figure}

Left panel of Fig.~\ref{fig:eta_s_T} shows variation of the ratio of shear viscosity ($\eta$) 
to entropy density ($s$) with the temperature for fixed value of the 
relaxation time $\tau = 1$ fm. The $\eta/s$ is calculated for different 
sizes of the system. Similar to $\eta$,
volume dependence is observed significantly
in this ratio especially at low temperature. 
Interestingly, the $\eta/s$ increases when we decrease the $R$.
It is also demonstrated by right panel of Fig.~\ref{fig:eta_s_T},
which discloses basically the fractional change in $\eta/s$ at different values of $R$
with respect the infinite matter result. 
There is opposite $R$ dependency between $\eta$ and $\eta/s$.
The $R$ dependence of entropy density $s$, shown in Fig.~\ref{fig:s_sSB_T},
will help us to understand the difference. In y-axis of Fig.~\ref{fig:s_sSB_T},
the entropy density is normalized by its Stefan-Boltzmann (SB) limit ($s_{SB}$) and 
a dimensionless quantity $s/s_{SB}$ is presented.
For 3 flavor quarks, the SB limit of entropy density is $s_{SB}=\frac{19\pi^2}{9}T^3$.
We notice that $s$ is decreasing
as $R$ decreases and in $\eta/s$, rate of decreasing for $s$ becomes dominant
over that for $\eta$. Therefore, $\eta/s$ ultimately increases as $R$ decreases.
LQCD data of $s/s_{SB}$ from the WB group~\cite{Borsanyi:2013bia} (triangles)
and Hot QCD group~\cite{Bazavov:2014pvz} (circles) are in well agreement with 
our estimations from the HRG model for the temperature range studied.
The straight horizontal dotted line in the left panel of Fig.~\ref{fig:eta_s_T} denotes the
lower bound of $\eta/s$, known as KSS bound~\cite{KSS}. So, our investigation
suggests that finite system size effect of hadronic matter leads to a shift in
$\eta/s$ away from the KSS bound. The values of $\eta/s$ below KSS bound
should not be related with any violation as estimations are proportionally sensitive
with relaxation time, which is arbitrarily chosen as 1 fm for this (left panel) Fig.~\ref{fig:eta_s_T}. 
Reader should focus on the changes
of $\eta/s$ with $R$, as shown in the right panel of Fig.~\ref{fig:eta_s_T}, instead of its absolute value.
However, for absolute value case, one should calculate the relaxation time in microscopical way as
done in Refs.~\cite{Itakura,Dobado,Nicola,Weise1,SSS,Ghosh_piN,Gorenstein,HM,Kadam:2015xsa,Hostler},
where it decreases as a function of temperature. Therefore, at low temperature, we will get a
higher value of relaxation time, for which $\eta/s$ will not go below the KSS bound. The temperature range
below 100 MeV may not be relevant domain for the phenomenological point of view as the medium
freezes out below that temperature.

%
\begin{figure} 
\begin{center}
\includegraphics[width=0.5 \textwidth]{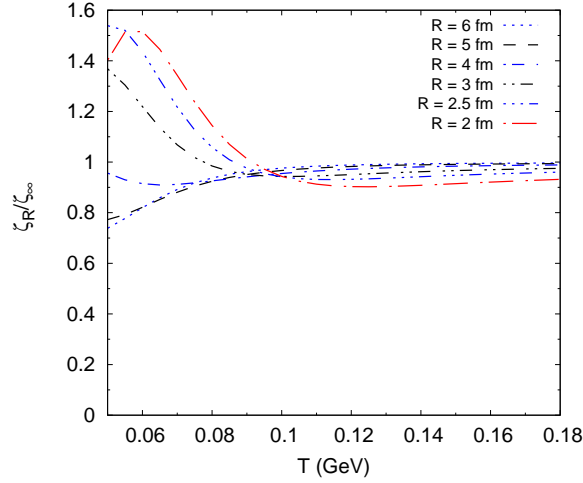}
\caption{(Color online) Same as Fig.~\ref{fig:eta_relative_T} but
for bulk viscosity of hadronic matter.} 
\label{fig:bulk_relative_T}
\end{center}
\end{figure}
\begin{figure} 
\begin{center}
\includegraphics[width=0.5 \textwidth]{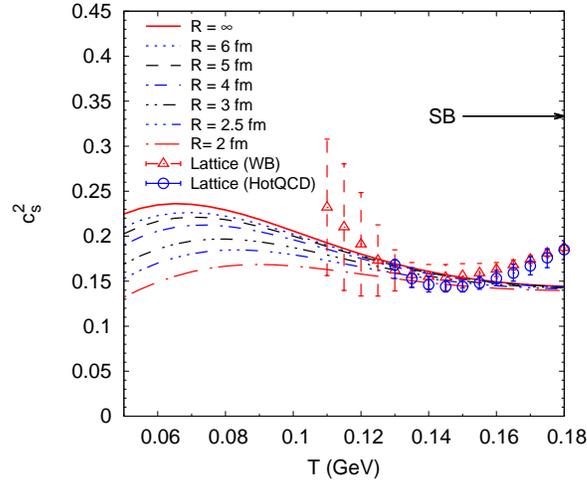}
\caption{(Color online) Square of speed of sound ($c_s^2$) versus $T$ for different values of $R$.
The Stefan-Boltzmann limit of $c_s^2$ is shown by the arrow. Corresponding 
LQCD data from the WB group~\cite{Borsanyi:2013bia} (triangles) and Hot QCD group~\cite{Bazavov:2014pvz} 
(circles) are also shown.} 
\label{fig:cs2_T}
\end{center}
\end{figure}
\begin{figure} 
\begin{center}
\includegraphics[width=0.42 \textwidth]{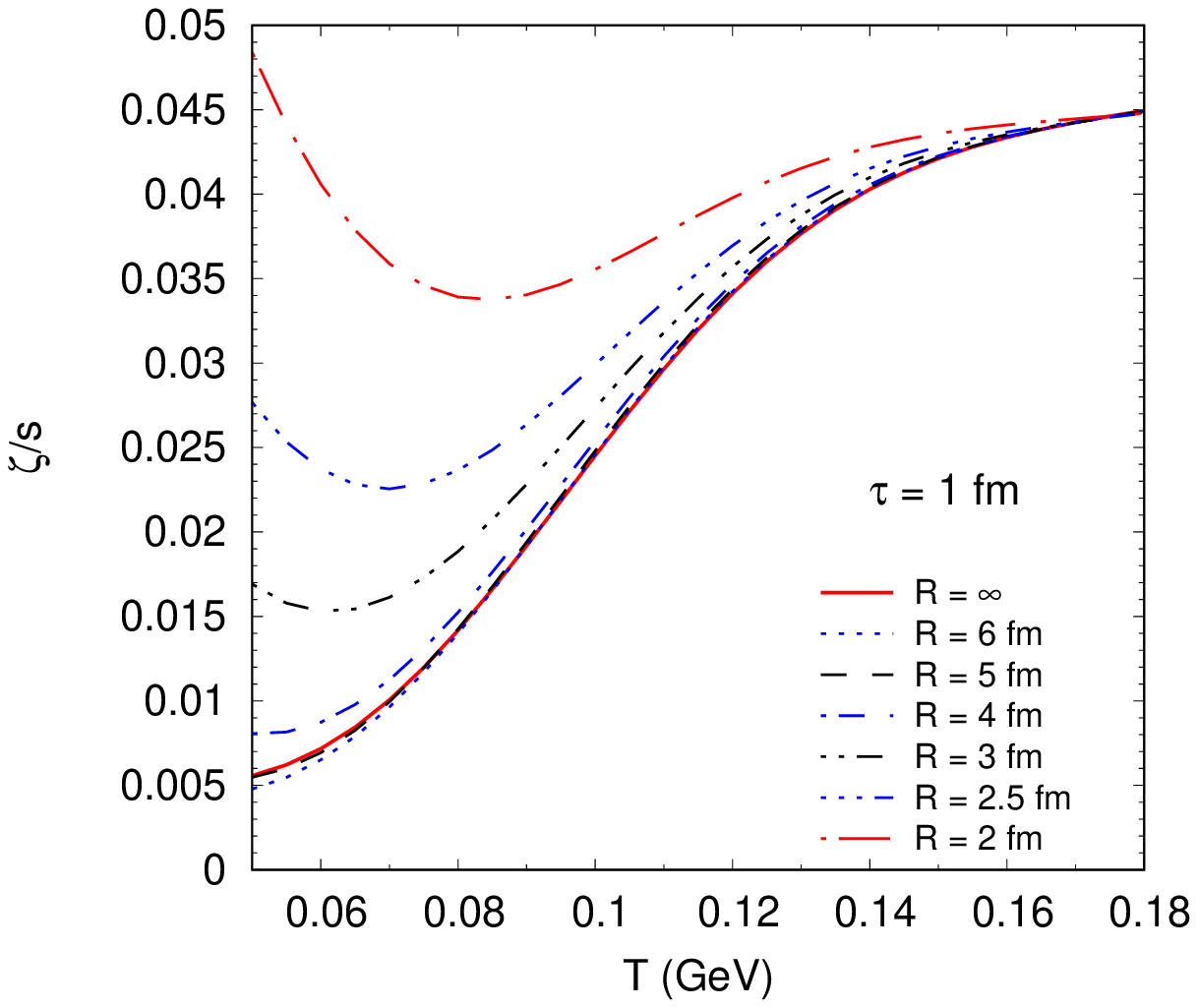}
\includegraphics[width=0.42 \textwidth]{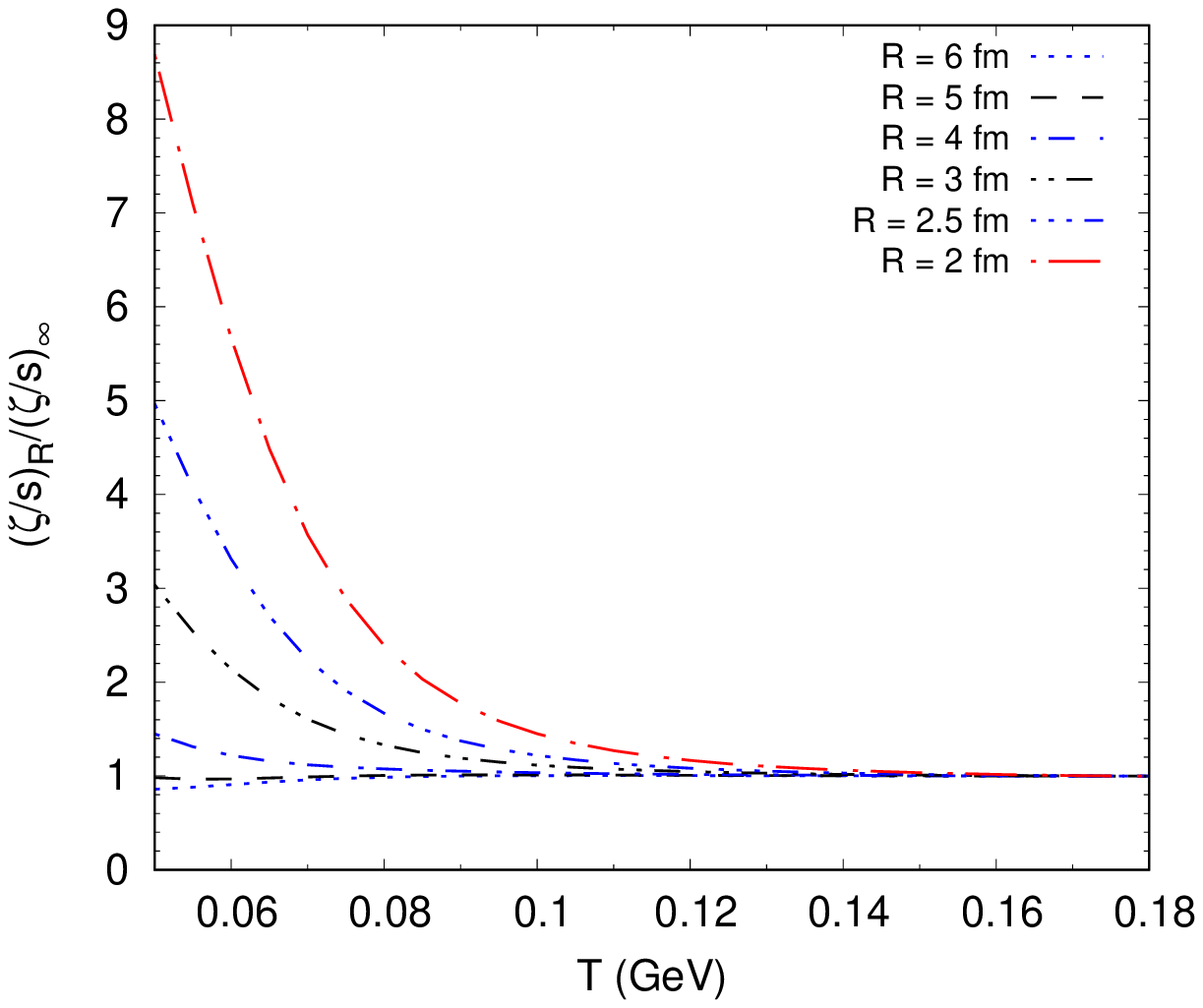}
\caption{(Color online) Left: The temperature dependence of bulk viscosity to entropy density 
ratio ($\zeta/s$) for different values of $R$. The calculations are done
for $\tau=1$ fm and $\mu=0$ MeV. Right: Fractional change of $\zeta/s$
at different values of $R$ with respect the infinite matter result.} 
\label{fig:bulk_s_T}
\end{center}
\end{figure}
\begin{figure} 
\begin{center}
\includegraphics[width=0.5 \textwidth]{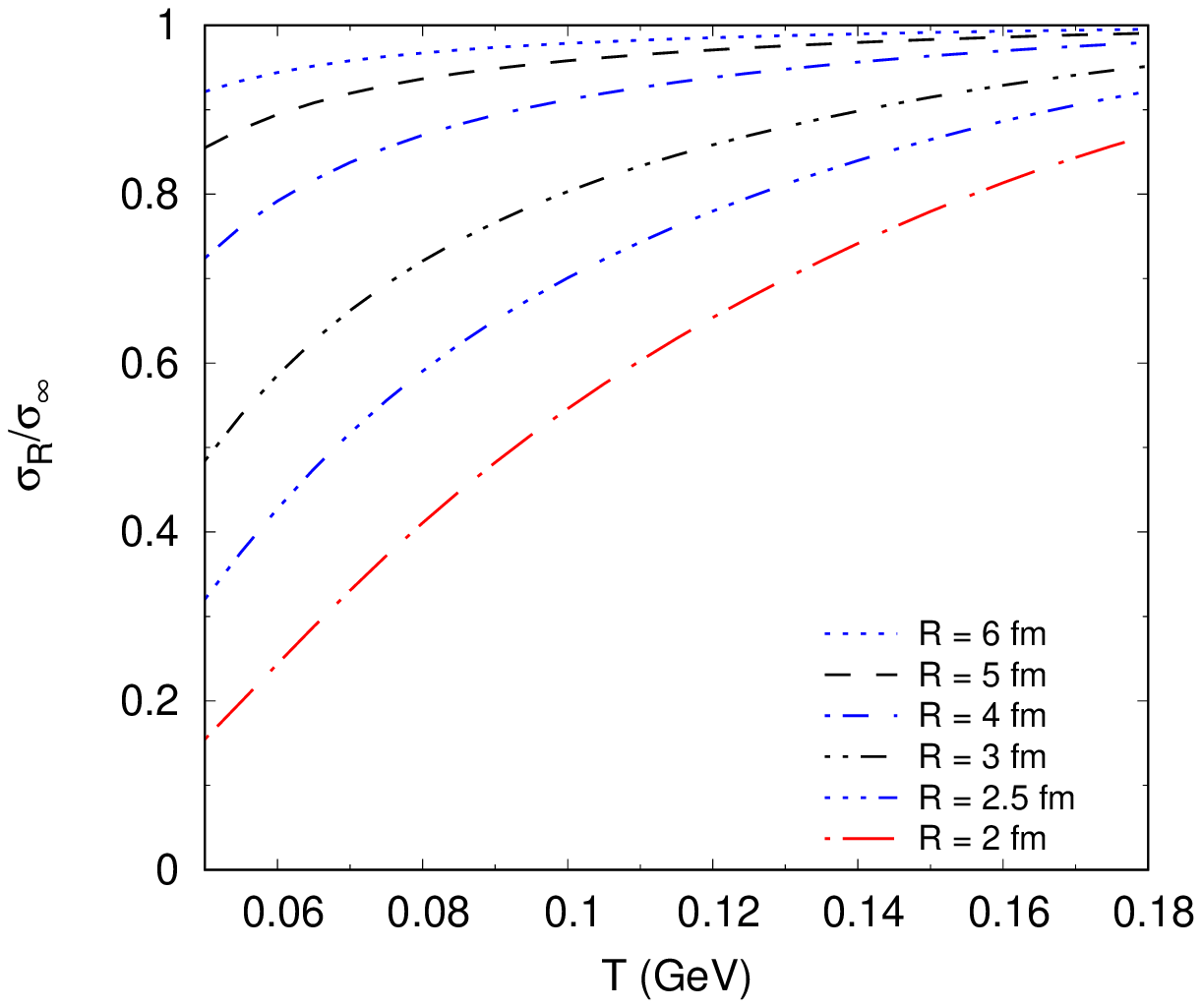}
\caption{Same as earlier Figs.~\ref{fig:eta_relative_T} and \ref{fig:bulk_relative_T},
for electrical conductivity of hadronic matter.} 
\label{fig:el_relative_T}
\end{center}
\end{figure}
%

The Fig. \ref{fig:bulk_relative_T} is same as Fig.~\ref{fig:eta_relative_T} but
for bulk viscosity of hadronic matter. 
It is observed that the ratio $\zeta_R/\zeta_\infty$ is more than unity at low temperature
for very small system sizes. 
If we look at Eq.~(\ref{eta_G}) for $\eta$ and Eq.~(\ref{zeta_Gmu0})
for $\zeta$, one can identify that the expression of $\zeta$ contain an
additional term 
\be
(1/3 -c_s^2)^2\vk^2 -c_s^2m_{B,M}^2~, 
\ee
which vanishes in the limits of $c_s^2\rightarrow 1/3$ and $m_{B,M}\rightarrow 0$.
At high temperature QCD, this limits hold and QCD matter reaches to a scale
independent or conformal symmetric situation, which can alternatively be realized
from the vanishing values of $\zeta$ for QCD matter at high temperature.
In this context, the bulk viscosity calculation in HRG model is trying to measure indirectly 
the breaking of this conformal symmetric nature of QCD matter in
the temperature range, where it is non-perturbative. 
The present investigation has tried to capture more delicate
issue - finite system size effect on this breaking of conformal symmetry
by studying the $R$ dependence of $\zeta$. 
The reason may be well understood from the Fig.~\ref{fig:cs2_T}, which shows temperature dependence of
square of speed of sound ($c_s^2$) for different values of $R$.
We notice that $c_s^2$ is suppressed at low temperature due to finite system size effect.
The conformal symmetry breaking term $(1/3-c_s^2)$ is enhanced due to this suppression 
in $c_s^2$. Therefore, the ratio $\zeta_R/\zeta_\infty$ becomes more than unity 
in the low temperature region.
LQCD data of $c_s^2$ from the WB group~\cite{Borsanyi:2013bia} (triangles)
and Hot QCD group~\cite{Bazavov:2014pvz} (circles) are included in Fig.~\ref{fig:cs2_T}.
Our estimations of $c_s^2$ in HRG model are in reasonable agreement with LQCD calculations 
in the temperature range studied.
At high temperature domain, where $c_s^2$ becomes more or less volume independent, $\zeta$
will decreases for smaller $R$ (similar to $R$ dependence of $\eta$).

The left panel of Fig.~\ref{fig:bulk_s_T} shows the $\zeta/s$ vs $T$
for different values of $R$, where we have again used $\tau=1$ fm for rough estimation.
For an infinite system, $\zeta/s$
increases with increase of temperature, but for a small system size, the $\zeta/s$ initially
decreases with increase of temperature and then at higher temperature it
increases slowly. The fractional change in $\zeta/s$ for different values of $R$ 
with respect its values for infinite matter case is shown in the right panel 
of Fig.~\ref{fig:bulk_s_T}. By taking a closer look at on the left and right panels of
Fig.~\ref{fig:bulk_s_T}, then we will find two different $R$-$T$ domains,
showing opposite behavior in $\zeta/s$. 
When we go from $R=6$ fm to $4$ fm, $\zeta/s$ decreases but it increases
for further reduction of $R$ ($4-2$ fm). Similarly $\zeta/s$ decreases at
low $T$ and increases at high $T$. Hence, we can visualize two domains in 
$R$-$T$ plane and their boundary, through which the $\zeta/s(R,T)$ face 
the transformation from its increasing to decreasing trends. The fact may
not be well revealed unless we visualize or zoom in the graphs. 
If we notice the expression of bulk viscosity, given in Eq.~(\ref{zeta_Gmu0}), 
we can find the reason of two different domain in $R$-$T$ plane. In one hand,
$\zeta$ is reduced because of the momentum cut-off by shrinking the phase space.
On the other hand, it is enhanced because the conformal breaking term $(1/3 -c_s^2)$
increases when we decrease the $R$. Hence, these two opposite effect are basically 
acting on $\zeta$ and as a net outcome, $\zeta$ (as well as $\zeta/s$) is displaying
opposite behavior in two different zones in $R$-$T$ plane.
%

Figure \ref{fig:el_relative_T} shows the results for the electrical conductivity of hadronic matter.
The presentation follows the same pattern as we have taken in 
earlier Figs.~\ref{fig:eta_relative_T} and \ref{fig:bulk_relative_T}.
With respect to other transport coefficients like $\eta$ and $\zeta$, the
electrical conductivity $\sigma$ is quite sensitive to the system size effect even at a higher
temperature. For example, at $T=160$ MeV, suppression in $\eta$, $\zeta$ and $\sigma$
for respectively $R=2$ fm are $1\%$, $8\%$ and $20\%$ with respect to $R=\infty$ case.
%
%
%
\begin{figure} 
\begin{center}
\includegraphics[width=1 \textwidth]{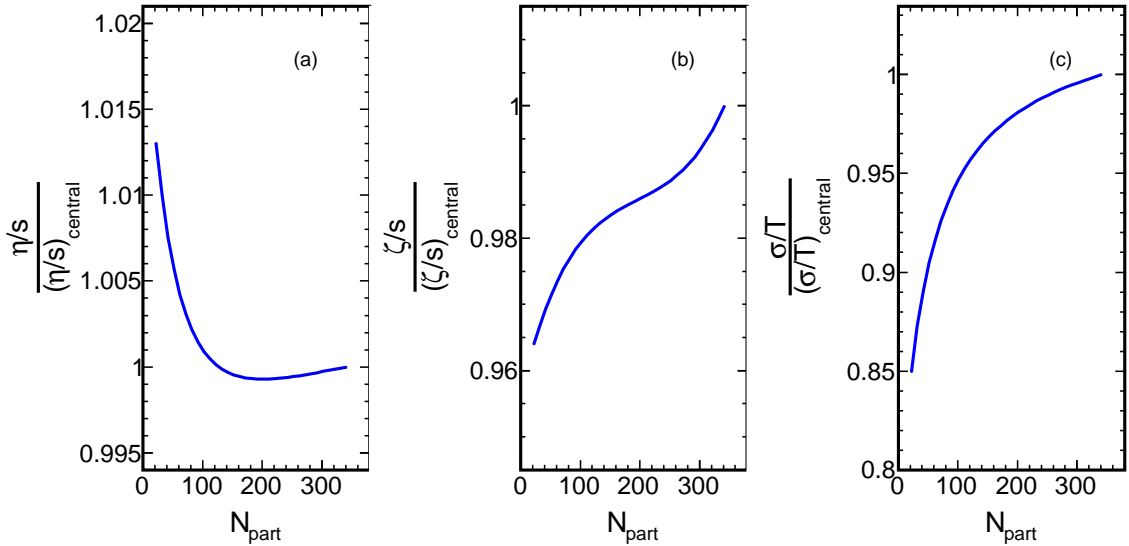}
\caption{Fractional change in $\eta/s$ (a), $\zeta/s$ (b) and $\sigma/T$ (c)
with respect to their values in most central collision.} 
\label{tr_coeff_npart}
\end{center}
\end{figure}

\begin{figure} 
\begin{center}
\includegraphics[width=1 \textwidth]{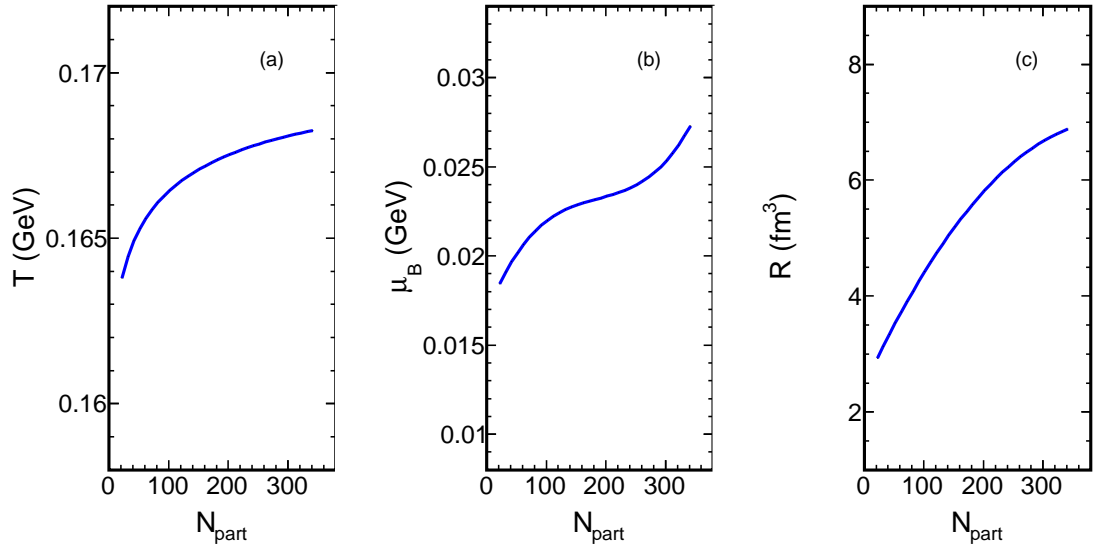}
\caption{$N_{\rm part}$ dependence of temperature $T$ (a), baryon chemical 
potential $\mu_B$ (b) and radius $R$ (c) of the medium at freeze-out point, 
taken from Ref.~\cite{STAR_17}.} 
\label{R_npart}
\end{center}
\end{figure}

Now let us come to the perspectives of our results in heavy ion phenomenology.
The expanding matter, produced in heavy ion collision experiments,
can be well described by dissipative hydrodynamic simulations, where
the transport coefficients like shear and bulk viscosities are used as input
parameters. During the expansion, the volume of the medium increases and temperature
decreases with time and at freeze-out temperature, the medium looses its many body
identity. 
This freeze-out system size can only be measured in experiment but in the initial
stages, this size of the system can be smaller. 
Our present investigation shows that the values of $\eta/s$ and $\zeta/s$ 
can be changed at different system sizes, which are less than 6 fm (approx). 
So it suggests to consider size dependent (along with temperature dependent) 
$\eta/s$ and $\zeta/s$ during the complete evolution. In most central collision,
the freeze-out size is too large ($\sim 7-8$ fm) to get any volume effect in $\eta/s$ and $\zeta/s$
but they can have volume effect in earlier stages of hydrodynamical evolution. For non-central
collision, the freeze-out size becomes smaller. So one can expect different values
of transport coefficients for different centrality or average number of 
participants ($N_{\rm part}$), which is demonstrated in
Figure \ref{tr_coeff_npart}.
It shows variation of transport coefficients
with respect to the $N_{\rm part}$
for Au+Au collision at RHIC energy  $\sqrt{S_{NN}} = 200$ GeV.
Centrality dependence of chemical freeze-out parameters like 
$T, \mu_B$ and fireball size $R$ are taken from the data of Table VIII, 
given in Ref.~\cite{STAR_17}. After parameterizing those data the values
are shown in Fig.~\ref{R_npart}.
These data of $T$, $\mu_B$ and $R$ are obtained by fitting the experimental data
of hadronic yields.
As we go from central to non-central collisions, the $N_{\rm part}$ decreases.
This centrality and $N_{\rm part}$ can be linked with standard Glauber model.
The experimental observations~\cite{STAR_17} tell that
freeze-out size of the medium reduces with decrease of $N_{\rm part}$. 
From the Fig.~\ref{R_npart}, we can roughly relate the zones $R=7-3$ fm, 
$\mu_B=0.027-0.019$ GeV and $T=0.168-0.164$ GeV for the $N_{\rm part}=350-40$.
Now, smaller $N_{\rm part}$ corresponds
to smaller $R$, where all of the transport coefficients $\eta$, $\zeta$ and $\sigma$ 
become smaller, as we noticed in Figs~\ref{fig:eta_relative_T}, \ref{fig:bulk_relative_T}, 
\ref{fig:el_relative_T}. Hence, all of the transport coefficients will
be decreased when we decrease the $N_{\rm part}$. Same trend can be expected 
for dimensionless quantity $\sigma/T$, as we see in Fig.~\ref{tr_coeff_npart}(c).
For dimensionless quantities $\eta/s$ and $\zeta/s$, the picture will be little complicated
because of additional impact of $s(R,T,\mu_B)$.
For $\zeta/s$ at the particular zone of $T$-$\mu_B$-$R$ will decrease as $N_{\rm part}$ decreases,
which we get in Fig.~\ref{tr_coeff_npart}(a). We can get a rough idea of this trend from
Fig.~\ref{fig:bulk_s_T}, where $\zeta/s$ can reduce as we move from $R=6$ to $4$ fm for 
$T\sim 0.164-0.168$ GeV. Although, there is a possibility of enhanced $\zeta/s$ at low 
$R$ and low $T$ domain, which is not relevant in the phenomenological direction.
From Fig.~\ref{fig:eta_s_T}, we have already noticed that $\eta/s$ is enhanced at
low $R$, from where one can relate the increasing nature of $\eta/s$ at low $N_{\rm part}$,
which we see in Fig.~\ref{tr_coeff_npart}(a). This qualitative $N_{\rm part}$ dependence 
of $\eta/s$ is in well agreement with the outcome of hydrodynamical simulation, where
$\eta/s$ is entered as an input parameter.
Ref.~\cite{Roy:2012jb} has studied the centrality or $N_{\rm part}$ dependence of
invariant yield and elliptic flow of charged hadrons as a
function of transverse momentum. By taking different guess values of $\eta/s$
in their hydrodynamical simulation,
they have attempted to match the experimental data of PHENIX Collaboration~\cite{PHENIX_pt,PHENIX_v2} 
and they found the experimental data
prefers higher values of $\eta/s$ as we go from central to peripheral
collisions. So we observe a similar qualitative trend of $N_{\rm part}$ dependence of
$\eta/s$ from the direction of hydrodynamical simulation~\cite{Roy:2012jb} as well as
from the microscopic calculation of present investigation.

In present work, 
we have started with simple picture of finite size consideration to build the $R$
dependent of $\eta/s$ and other transport coefficients. In future, we have a plan to
consider other possible aspect of finite size picture, which may rigorously change
the size dependence of different transport coefficients. However, a new phenomenological
message from this present work is that the $\eta/s$ (as well as other coefficients)
can depend on system size along with the other medium parameters $T$, $\mu_B$. Our
results suggest to consider an explicit $R$ (as well as $T$, $\mu_B$) dependent 
$\eta/s$ in hydrodynamical simulation, which may help to match the experimental data
more precisely. We also plan to explore this phenomenology in future.

Another important quantity is relaxation time, which decide the numerical
strength of transport coefficients. In the present work, 
as discuss in most of the cases the ratio of transport coefficients with and without finite 
size effects, the contribution from relaxation time cancels.
This is true if the relaxation time is a constant quantity for the system.
A microscopic calculation may provide us a momentum and temperature
dependent relaxation time, due to which our results may be changed slightly, which may be very
interesting to study in near future.

%
\section{Summary and Perspectives} 
\label{sec:summ}

In summary, we have tried to investigate the finite system size effect
on different transport coefficients of hadronic matter.
We have used an ideal HRG model to describe the thermodynamic behavior
of hadronic medium constituents. We have adopted the simplest possible
idea of a finite lower momentum cut-off to study the 
finite system size effect on the transport coefficients of the medium. Using standard expressions
of transport coefficients like shear viscosity $\eta$, bulk viscosity $\zeta$
and electrical conductivity $\sigma$, we have calculated them as a function
of temperature $T$ for various system sizes $R$.
We observe a significant finite system size effect on these transport
coefficients at low temperatures.
However, this effect reduces as we go
to higher temperatures. The values of shear viscosity and electrical conductivity
decrease due to finite system size effect. This is also true for bulk viscosity
but at high temperature. At low temperature domain, when we decrease the $R$, we get a 
suppression in speed of sound and therefore an enhancement in  $\zeta$ is observed. 
%
%
The entropy density also gets suppressed due to finite system size effect. The 
competing dependence of $\eta$ and $\zeta$ with $s$ on system size gets reflected in the dimensionless
ratios $\eta/s$ and $\zeta/s$. 
For smaller values of $R$, the $\eta/s$ becomes larger. In case of $\zeta/s$, suppression
is observed when $R$ will be reduced from $6$ fm to $4$ fm but for further reduction, an
enhancement is noticed. 

In phenomenological point of view, we have also seen the centrality or average number of 
participants $N_{\rm part}$ dependence of $\eta/s$, $\zeta/s$ and $\sigma/T$. In heavy ion collisions experiment,
the picture from central to peripheral collisions is linked with increasing centrality and decreasing
$N_{\rm part}$ and $R$.
We notice that $\zeta/s$ and $\sigma/T$ become smaller but $\eta/s$ is larger at smaller $N_{\rm part}$.
From the direction hydrodynamical simulations, a larger $\eta/s$ is required as input parameter
to describe lower $N_{\rm part}$ (higher centrality) experimental data. Hence, we notice that the macroscopic
hydrodynamical simulation and our microscopic calculations both indicate a similar trend of
$N_{\rm part}$ dependence in $\eta/s$.  Our
results also suggest to consider an explicit size dependent 
$\eta/s$ in hydrodynamical simulation during the entire expansion.
In near future, we are planning to incorporate other possible rigorous tools of finite size effect
in the microscopic calculations of transport coefficients, while in long-term plan, we are also
planning for proper implementation in macroscopic evolution picture.

{\bf Acknowledgment:} 
SG is partially supported from  University Grant Commission (UGC) 
Dr. D. S. Kothari Post Doctoral Fellowship (India) under
grant No. F.4-2/2006 (BSR)/PH/15-16/0060. SS and BM are supported by DAE
and DST, which is greatly acknowledged. We thank to Victor Roy for useful
discussions. SG acknowledge
WHEPP-2017 for getting some fruitful discussions.

\end{document}